\documentclass[acmsmall,table]{acmart}

\usepackage[inline]{enumitem}
\usepackage{subcaption}

\usepackage{multirow}
\usepackage{colortbl}
\usepackage{tabularx}
\usepackage{ltablex}
\usepackage{geometry}

\newcommand{\bs}[1]{\textcolor{purple}{\textbf{*Bahar*}: #1}}

\newcommand{\skj}[1]{\textcolor{blue}{\textbf{*Sujay*}: #1}}
\newcommand{\rw}[1]{\textcolor{red}{\textbf{*Ryen*}: #1}}

\newcommand{\rvs}[1]{\textcolor{black}{#1}} 
\newcommand{\cmr}[1]{\textcolor{black}{#1}} 

\newenvironment{myquote}[1]%
  {\list{}{\leftmargin=#1\rightmargin=#1}\item[]}%
  {\endlist}

\AtBeginDocument{%
  \providecommand\BibTeX{{%
    \normalfont B\kern-0.5em{\scshape i\kern-0.25em b}\kern-0.8em\TeX}}}

\setcopyright{acmcopyright}
\acmJournal{PACMHCI}
\acmYear{2020} \acmVolume{4} \acmNumber{CSCW} \acmArticle{1} \acmMonth{10}

\begin{document}

\title{Characterizing Stage-Aware Writing Assistance in Collaborative Document Authoring}

\author{Bahareh Sarrafzadeh}
\authornote{Work done while a PhD candidate at University of Waterloo and interning at Microsoft Research}
\affiliation{%
  \institution{Microsoft}
  \city{Redmond}
  \state{WA}
  \country{USA}
}
\email{bsarrafz@uwaterloo.ca}

\author{Sujay Kumar Jauhar}
\affiliation{%
  \institution{Microsoft Research AI}
  \city{Redmond}
  \state{WA}
  \country{USA}
}
\email{sjauhar@microsoft.com}

\author{Michael Gamon}
\affiliation{%
  \institution{Microsoft Research AI}
  \city{Redmond}
  \state{WA}
  \country{USA}
}
\email{mgamon@microsoft.com}

\author{Edward Lank}
\affiliation{%
  \institution{University of Waterloo}
  \city{Waterloo}
  \state{ON}
  \country{Canada}
}
\email{lank@uwaterloo.ca}

\author{Ryen W. White}
\affiliation{%
  \institution{Microsoft Research AI}
  \city{Redmond}
  \state{WA}
  \country{USA}
}
\email{ryenw@microsoft.com}

\renewcommand{\shortauthors}{Sarrafzadeh, et al.}

\begin{abstract}
Writing is a complex non-linear process that begins with a mental model of intent, and progresses through an outline of ideas, to words on paper (and their subsequent refinement).  
\rvs{Despite past research in understanding writing, Web-scale consumer and enterprise collaborative digital writing environments are yet to greatly benefit}
from intelligent systems that understand the stages of document evolution, providing opportune assistance based on authors' 
situated actions and context.  
\rvs{In this paper, we present three studies
that explore temporal stages of document authoring.} 
We first survey information workers at a large technology company about their writing habits and preferences, concluding that writers do in fact conceptually progress through several distinct phases while authoring documents.  
\rvs{We also explore, qualitatively, how writing stages are linked to document lifespan.}
We supplement these qualitative findings with an analysis of the longitudinal user interaction logs of a popular digital writing platform over several million documents. Finally, as a first step towards facilitating an intelligent digital writing assistant, we conduct a preliminary investigation into the utility of user interaction log data for predicting the temporal stage of a document. Our results support the benefit of tools tailored to writing stages, identify primary tasks associated with these stages, and show that it is possible to predict stages from anonymous interaction logs. Together, these results argue for the benefit and feasibility of more tailored digital writing assistance.

\end{abstract}

\keywords{Writing; Collaborative Writing; Intelligent writing assistance; Workplace; Productivity}

\maketitle

\section{Introduction}

When authoring a document, a writer does not go from a blank page to a fully completed text in a linear fashion. Rather they may go through several overlapping, sometimes cyclical, \emph{stages} \cite{rohman1965,flower1984} of planning, outlining, drafting, revising, soliciting feedback, and copy-editing to crystallize their thoughts into words. Progressing through these stages, the author might have very different mental models of, or approaches towards, their goals and writing needs at any given point in time. For example, in drafting mode, their goal might be to simply write down thoughts, without worrying too much about spelling or grammar; whereas in copy-editing mode, they might want to go back and correct their typographical or grammatical errors.

While digital writing environments include a number of tools (such as grammar- and spell-checkers, anchored comments, change tracking, etc.) to support authors as they progress through the stages of writing a document, the surfacing of these tools through the user experience can sometimes be inopportune, causing the writer to break their focus. Despite the widespread use of context-aware intelligent systems in many domains such as conversational search \cite{radlinski2017}, travel planners \cite{borras2014}, and calendar-aware email recommendation \cite{zhao2018}, modern writing tools often ignore the notion of writers' ``situated actions'' \cite{suchman1987} beyond the most recent changes to a document's text. 

While better individual writing support may benefit any individual author, understanding document evolution becomes even more important, we would argue, as writing moves online, primarily because one of the primary benefits of these online platforms is their support for collaborative document authoring. 
\citet{greer2016} note that, if writing is decomposed into distinct rhetorical phases, this phased writing process helps structure writing, and can guide collaborators co-creating documents to appropriate subtasks during different phases of document creation \cite{kellogg1988, teevan2016}. Understanding -- and depicting 
-- document phases can also help new collaborators get up-to-speed, while supporting efforts at proper collaboration etiquette\cite{gutwin2002, endsley2017toward}.

While there is broad awareness that many documents are created collaboratively \cite{teevan2016}, research on document evolution and stages of writing has been restricted to laboratory studies or theoretical investigations. Previous efforts include the analysis of writing captured on video \cite{xu2014}, thinking-aloud methods \cite{flower1981}, and retrospective interviews \cite{posner1992, birnholtz2012}.
Most of these previous approaches focus on specific kinds of writing (e.g., essay writing in classroom \cite{yim2017}), require that documents be of a certain quality (e.g., project reports \cite{olson2017} or featured articles in Wikipedia \cite{yang2017}), or are limited in scope -- and are therefore ill-suited to supporting a general-purpose Web-scale collaborative digital writing environment used by millions of users for heterogeneous document authoring.



\rvs{In this paper, 
we probe the question of whether and how authors divide their writing into stages and the link between authors' actions and the temporal lifespan of a document they create.} 
\cmr{While document lifespan can, in theory, be infinite, here we focus on the "active" lifespan of a document, i.e. the portion of the document's existence when it is being actively modified. We explore whether authors of collaboratively written documents perceive writing stages \cite{rohman1965,flower1984} and whether these stages can be discriminated via differences in user action over different periods of a document's (active) lifespan.}
To address this research goal, we leverage a mixed-methods approach \cite{creswell13} 
that combines data from three studies: a qualitative study \rvs{that explores the link between writing tasks and document lifespan}, a quantitative study that analyzes user behaviors over document lifetimes, and a preliminary predictive analysis that examines whether 
\cmr{different stages within a collaboratively authored document's lifespan} can be inferred from user interaction log data.  

\rvs{Our qualitative study surveys 183 information workers at a large technology company about their writing habits and finds that authors frequently collaborate around document creation and writing, and that authors have mental models of their writing goals, and these broadly correspond to \rvs{temporal} stages of a document's \rvs{lifespan}.  Authors also note that different authoring stages are associated with different types of activities on the writing canvas, and they perceive the benefits of \rvs{stage-aware writing} assistance that supports them in their situated actions and through the stages of their writing, particularly for collaborative writing tasks. They also offer insights into the types of support needed at different stages.}
\rvs{To validate the link between document lifespan and different writing activities}, we perform a follow-on quantitative analysis on the user interaction logs of several million documents authored on a popular Web-scale digital writing platform. Our quantitative log analysis confirms that writers distribute their writing activities differently over temporal stages of document \rvs{lifespan} both in terms of magnitude and types of these activities. The triangulation of our qualitative insights on authors' perceptions and user actions provides a better understanding of how authors engage with the writing task at different stages and serves as a guide to design of intelligent writing assistants that incorporate stage-appropriate assistance for writing activities.
Finally, as a first step towards demonstrating the feasibility of stage-appropriate writing assistance tools, we model the temporal stage of document evolution using the interaction logs from our quantitative study. Our results indicate that with a statistically significant margin, log-based interactions can be leveraged to predict \rvs{these stages of document evolution} more accurately than a strong baseline.

\rvs{One implication of our qualitative survey was that, while our log analyses indicate an ability to predict document stage, survey respondents were almost evenly split in their desire for automatic inference of document lifespan, primarily due to past negative experiences with intelligent writing assistants (e.g. Clippy). In our discussion, we revisit this aspect of our work, and highlight that document-lifespan inferences can be leveraged in many ways: as intelligent tools with passive user involvement; as defaults with an ability to over-ride; as a mixed initiative adaptation \cite{Bunt2007, horvitz1999}; or as an awareness mechanism to support more effective collaboration \cite{gutwin2002}.}

\section{Related Work}

There are several areas of research related to stages of document authoring and intelligent writing assistance. We review them in what follows and highlight differences to our work.
\subsection{Writing Process}
A number of papers have investigated the process of writing, both as an individual effort and as a collaborative endeavor.
Some studies focus largely on the text itself, characterizing writing as an ``outline-draft-edit'' cycle. These works assume that ``meaning'' is first assembled in the author's mind and then articulated in text \cite{mitchell1996}.
Other studies adopt the \textit{process} view of writing, basing their research on cognitive process models \cite{freedman1987} for problem solving. Broadly this involves sequences of activities, 
such as setting goals, planning, organizing, transcribing, and editing.
Fitzgerald \cite{fitzgerald92} presents a comprehensive overview of early approaches to the study of writing. She identifies three models of writing: the Stages Model \cite{rohman1965}, the Problem-solving Model \cite{flower1984}, and the Social Interaction Model \cite{nystrand1989}. 
While there is no single comprehensive theory of the writing process \cite{greer2016}, the Flower and Hayes model \cite{flower1981} is one of the more widely accepted theories. It considers writing as a series of cognitive processes.
Through a think-aloud study they showed that the writing process involves three main sub-processes: planning, translating, and reviewing.
Models of writing processes have been evaluated observationally in different ways, 
using -- for example -- video analysis \cite{xu2014}, think-aloud methods \cite{flower1981}, and retrospective interviews \cite{posner1992}, mostly conducted at small-scale and in controlled settings. There has also been some effort on using keystroke logging to measure writing processes during typing \cite{conijn2019, tillema2011, steck2018}.

Collaborative writing, on the other hand, builds on single author writing by involving multiple people, thus increasing the complexity of the writing process \cite{galegher1994}. Although collaborative writing is a group effort, many activities are often divided and conducted on an individual basis \cite{teevan2016, tammaro1997, lowry2004}. 
One of the reasons for the amplified complexity is the increased need to coordinate between multiple viewpoints and work efforts \cite{baecker1995}, and the need to establish consensus \cite{galegher1994, gutwin2002}. 
\rvs{Over the last three decades different HCI and CSCW researchers have focused on people's various practices in collaborative writing and how tools can support these practices. While some earlier works have largely involved improving technical capabilities of writing tools, such as updating revision histories \cite{ignat2008} or users' permissions or access controls \cite{ellis1991groupware, olson1993groupwork}, a different body of more recent work explores the adoption of collaborative writing tools long after they were developed and suggests that these tools or their collaborative capabilities are not necessarily used in practice \cite{noel2004, wang2017, wang2019, d2018spacetime}.
A range of social \cite{birnholtz2013}, personal \cite{d2018spacetime} and privacy \cite{wang2017} factors remained important barriers to the adoption of these tools. 
The main takeaway from this line of work is that, while many technological advances in designing CSCW tools are supported by past research and are expected to provide utility for the users, understanding users' attitudes towards and perception of these advanced functionalities should be at the heart of designing these new collaborative writing tools. Similarly, our qualitative study also provides insights regarding the acceptance of stage-aware tools and intelligent assistants that aim at optimizing users' writing practices, and whether fully automated solutions are appreciated by the users. 
}

\subsection{Decomposing the Writing Process}
In addition to research on the writing process itself, several previous studies have looked at the decomposition of the process into distinct stages.
The cognitive process model of writing \cite{flower1981} divides up document authoring into
distinctive and hierarchical phases, which enables writers to construct highly personal, idiosyncratic and complex workflows \cite{teevan2016}. The complexity of these workflows requires authors to coordinate multiple and often competing writing activities \cite{alamargot2001, alamargot2009} 
These processes compete for a writer's attention which can lead to cognitive overload \cite{kellogg1988, greer2016}.
Especially high demand processes, such as planning and reviewing, require significant levels of attention control \cite{kellogg1988, conijn2019}. Non-expert authors 
utilize strategies to reduce the cognitive load by \textit{decomposing} the writing task into smaller, more manageable subtasks and stages \cite{greer2016, graham2012, kellogg1988, limpo2013}. 
Effectively, this allows authors to scaffold their work, prompts thoughtful engagement and eases cognitive load \cite{teevan2016}.

\subsection{Revision Research}
Our work also relates to previous research on how revisions impact document evolution.
In the education domain, analysis of revisions has typically relied on manual observation of students' essay writing practices in classrooms, and coding of writing behavior, as well as grading the textual products of revisions. The main motivation behind this line of work is to analyze the relationship between revision and either text quality \cite{sommers1994} or their didactic effect \cite{murray1980}. For example, \citet{faigley1981} show that expert writers revise in ways that are different from inexperienced writers.
Differences between revisions contain valuable information for modeling document quality or extracting users' expertise and can additionally support various natural language processing (NLP) tasks \cite{yang2017}. The availability of vast amounts of data in Wikipedia revision histories has enabled modeling and applications in areas such as, 
sentence compression \cite{yamangil2008}, lexical simplification \cite{yatskar2010}, information retrieval \cite{aji2010}, textual entailment recognition \cite{zanzotto2010}, language bias detection \cite{recasens2013}, and spelling error correction and paraphrasing \cite{zesch2012, max2010}.
These tools, however, do not incorporate any awareness of the stage of writing or the state of document lifespan to provide more timely assistance and minimize interruptions to the users' workflow. 

\subsection{Contextual Recommendations}
Broadly, the idea of stage-aware writing assistance also relates to research on contextual recommendation systems \cite{adomavicius2011, shi2014, hariri2013}. This work aims to leverage a user's context to improve the quality of personalized services. For example, previous approaches have tailored recommendations by integrating user context in conversational search \cite{radlinski2017}, travel planners \cite{borras2014}, intelligent email reminders \cite{sarrafzadeh2019}, and calendar-aware email recommendation \cite{zhao2018}. Relatively little prior research has focused on contextual modeling for writing assistance, although \citet{horvitz1998} did explore using Bayesian inference to understand writers' intents and goals given their actions and queries.

\rvs{Our work extends past research in two important ways. First, while past work examines the writing process \cite{rohman1965, freedman1987, flower1984} and decomposing this process \cite{teevan2016}, the link between stages of writing and document lifespan, particularly for multi-authored documents, is unclear. While it seems reasonable that writing stages would correlate with document lifespan \cite{rohman1965}, understanding authors' perspectives on the link between writing activities and document lifespan has not been clearly established in past research. Second, past document-based studies have often included text and word-level features of documents to understand writing stages. Our analysis is both larger in scale (millions versus thousands of documents) and leverages anonymized user interaction logs that do not include character or word-level features of a document.}


\section{Stage-Aware Writing Assistance}
\label{QualitativeStudy}
\rvs{Our work is motivated by a desire to understand, cognitively, how and whether authors perceive that there are stages in their writing process.  Alongside this, we wished to explore how writing stages are linked to document lifespan and whether tools linked to document lifespan might better support collaborative writing tasks.}  

\subsection{Survey Description and Analysis}
\label{Survey}
To investigate whether providing writing assistance that is tailored to the current stage of a document under development is beneficial to the writers of that document, we need to better characterize two different aspects of \textit{stage-aware writing assistance}: 
\begin{enumerate}
    \item \textbf{Characterizing the stage of a document:} How are the stages of document evolution perceived by the writers?
    \item \textbf{Characterizing stage-appropriate writing assistance:} \\What writing assistance is beneficial at different stages of document evolution?
\end{enumerate}

To elicit people's perspectives on writing stages and whether stage-appropriate assistance might be beneficial, we distributed a survey within a large technology company. The survey included both closed and open-ended questions. We collected 183 completed responses to this survey (response rate: 10\%, completion rate: 74\%). 

Our survey started with a section on demographics including job roles, number of current projects, and team size. The next section involved different questions regarding writing documents at the workplace. These questions helped us verify whether writing documents at the workplace is a common practice, what tools are frequently used for writing documents, and how often these documents involve other collaborators.
Finally, the main segment of our survey asked participants to reflect on their perception of the utility of an intelligent writing assistant and how stage-aware writing assistance should be realized.

Respondents came from a diverse set of roles ranging from product managers and software developers to sales people and administrative assistants, the majority had been working at the company for more than 5 years (63\%).  Multi-tasking on different projects was a common trend among them, with 75\% working on at least 3 projects in parallel at the time of completing the survey.

\subsubsection{Analysis.}
The responses to closed-ended questions were aggregated to provide general statistics on participant demographics and writing habits.  
\rvs{The responses to open-ended questions were analyzed using affinity diagramming. We use the K-J method for affinity diagramming (i.e. paper-note-quote based affinity diagramming) 
\cite{harboe2015, holtzblatt1997}. As per Corbin and Strauss \cite{corbin1990grounded}, we separate our coding into initial open coding (19 cluster) and subsequent axial coding where we identify causal factors, connections, implications and other relationships between data to combine clusters into overall themes.  We found that saturation for our
qualitative data occurred approximately after 80 participants, and from this point no
new clusters of information were identified. However, we continued
to cluster responses for the remaining participants, particularly
attuned to data that might expand our clusters or add nuance to
our analysis. }

\subsection{Qualitative Findings} 
\label{QualitativeFindings}


We found that 87\% of respondents mentioned writing documents as a part of their work routine. They use a wide range of writing tools including Microsoft Word's Desktop client (89\%) and Online (Web-based) client (49\%),  Notepad / Notepad++ (19.3\%), OneNote (16\%), Overleaf / Sharelatex (11.1\%), Google Docs (8.2\%), and others.

Collaboration on these documents was a common theme, with the majority of participants (74\%) collaborating on at least 25\% of their documents.
A closer examination of the qualitative responses (discussed below) indicated that the need for tailored support in different stages of document authoring is associated with both Single Author Writing (SAW) and Collaborative Writing (CW) scenarios. 
\begin{myquote}{0.1in}
\textit{``in an ideal scenario I would just do some sort of content dump to the system and the system structures this text. So that I don't have to think about paragraphs, sections, where to put what etc. And then in later stages I want help with how to nicely formulate things. That's not only in a collaborative setting though.''} [P10]
\end{myquote}
To understand whether there is a need for different types of support at different stages of document evolution, 
the participants were asked \textit{``Should an intelligent writing assistant provide different types of support at different stages of writing in the document?''} with a binary Yes / No response type.

Over 70\% of participants expressed a need for different types of support at different stages of writing. 
\rvs{To elicit the rationales behind these responses, the participants were presented with a series of follow-up questions that provided an opportunity for them to elaborate on their responses and helped with decoupling our participants' attitudes towards stage-aware assistance versus intelligent writing tools.}

\rvs{Participants who responded Yes to the above question were prompted to elaborate on their response by the following three questions:
\begin{itemize} 
    \item[$\textbf{[Q1]}$] Why do you think different types of support are needed at different stages of document authoring? Can you think of some scenarios where you needed different types of support at different times?''
    \item[$\textbf{[Q2]}$] ``Should the AI infer your current phase of writing (e.g. drafting vs reviewing vs editing) or would you like to specify it yourself?''
    \begin{itemize}
        \item[$\textbf{[Q3-a]}$] $[$AI should detect it$]$ ``How do you expect the Intelligent Assistant to customize your experience based on your inferred stage?''
        \item[$\textbf{[Q3-b]}$] $[$I'd like to specify it$]$ ``Why do you prefer to be in charge of specifying your current phase of writing to the system?''
    \end{itemize}
\end{itemize}
Participants who responded No to the initial question were asked ``\textit{Why do you think different types of support are NOT needed at different stages of writing?}''}

\rvs{A closer examination of these responses to the follow-up questions indicated that while 70\% of all participants were in favor of stage-aware writing assistance they were mixed between a desire for the AI system to detect this stage (40\%), specifying the stage themselves (40\%) and Mixed-initiative (20\%).
Over 68\% of participants who responded No to the initial question as well as 65\% of participants who responded Yes to this question but mentioned they would like to specify the stage themselves cited bad experiences with the performance of AI systems and a lack of trust in the ability of intelligent systems to get this right as the main rationale behind their reluctance to the idea of intelligent stage-aware writing assistants.}
\rvs{Cumulatively, we find participants separated into three groups: approximately half of participants viewed some form of intelligent, stage-aware writing assistance as valuable; an additional 40\% were in favor of stage-aware writing tools, but wished to specify themselves, which tool sets to use; finally, 10\% of participants were indifferent.}

\rvs{In coding our responses to these questions, one theme that emerged from our data was a link between temporal stages of document authoring and temporal stages of writing.  While Teevan et al. \cite{teevan2016} break collaborative writing tasks down into microtasks to support the idiosyncratic nature of workflows, one aspect that has been unexplored by past work is this link between temporal stages of writing and temporal evolution of a document.} 
\rvs{To probe this issue in additional detail, 
the qualitative analysis of the follow up open-ended questions resulted in four broad themes, corresponding to different notions of the stage of a document as perceived by our participants and the utility of different types of writing assistance based on these stages.}

The following three subsections elaborate on different perceptions of a document's stage and their corresponding writing activities, which has implications for designing stage-aware writing assistants. The final subsection provides a different perspective on the utility of document stage awareness, which is essential for more effective collaborative writing scenarios.



\subsubsection{Temporal Stages of Document Authoring}
The Stages Model of Writing \cite{rohman1965} focuses on the evolution of the text as an external by-product of the writing where the main three stages of pre-writing, writing, and rewriting are separated in time. 
This notion of document stages resonated with many of our participants, who referred to them as early versus later stages of document writing. 
Qualitative analysis of the respondent's descriptions of these temporal stages revealed how they are conceptualized and what writing phases and activities are associated with each stage.
Later stages, on the other hand, are described as the time when most of the content has been added to the document and the writers can now focus on improving different aspects of the document during Reviewing, Copy-Editing and Finalizing phases.

\begin{myquote}{0.2in}
    \textit{``Document creation and document perfection are fundamentally different.  In my opinion, the best way to create a collaborative document is to get a rough draft made as quickly and dirtily as possible and then begin to improve it; different forms of assistance are required for each scenario.''} [P237]
\end{myquote}

At later stages the document is more ``mature'' with most of the content already included and structured. This quality has at least two different implications for what support can and should be provided at earlier versus later stages of writing. First, in later stages 
the system can provide more end to end and global feedback as opposed to local suggestions
(e.g., identifying missing topics to be included versus spelling errors to be fixed). Second, with more content in place and a longer history of edits, intelligent writing assistants can be more reliable in predicting what support is needed.  
\begin{myquote}{0.2in}
    \textit{``When I'm actively working on creating the document I want suggestions around better writing style, but focused on what I'm doing at that moment.  Once the document is more mature I'd want feedback on the end to end document, not a specific word here or there.''} [P106]\\
    \textit{``Intelligence should improve as the document matures.  The intent of the document, the subject matter, and the style of the document will become more clear as it evolves.''} [P78]
\end{myquote}

Finally, while supporting the activities that are associated with a certain stage was appreciated by survey respondents, untimely system support can be perceived as \textit{distracting}, \textit{annoying}, or \textit{unnecessary}.
In particular, minor fixes, proofing and grammar fixes, collaborative activities such as commenting and change tracking are perceived as unhelpful during early stages of document evolution.
\begin{myquote}{0.2in}
    \textit{``Worrying about complete sentences and grammar while brainstorming would slow things down considerably ... but necessary for a final release.''} [P227]\\
    \textit{``Initial document writing is mainly non-collaborative, whereas later it's a matter of tweaking and improving with the feedback of others.''} [P178]
\end{myquote}

The following three design implications are our main takeaways from the preceding analysis:
\begin{enumerate}
    \item Intelligent support should be provided based on the activities associated with the current stage of document evolution (i.e., early versus later stages).
    \item Authors are in a focused mode during earlier stages of writing and system interruptions should be minimized. 
    \item Authors can benefit from a higher level of control over the writing process in earlier stages with outlining, structuring, or searching functionalities that are provided on demand. During later stages, intelligent systems can be more proactive in providing feedback and suggestions to improve the overall document quality.
\end{enumerate}
\vspace{-1mm}

\subsubsection{Phases of Writing}
Different models of the writing process \cite{flower1984, lowry2004} state that writers go through different phases throughout the document authoring process. These phases are characterized as brainstorming, outlining, drafting, reviewing, revising and copy editing \cite{lowry2004} or more broadly as planning, translating, and reviewing \cite{flower1984}. 
Our participants' responses reveal that the idea of going through different phases during the process of writing documents resonates with them and that they perceive the stage of the document based on these phases of writing. In fact these phases were described in a variety of terms:
\textit{``from document creation to document perfection''} [P237], \textit{``starting from nothing phase''} [P98], \textit{``Rough draft vs final draft''} [P79], \textit{``Draft, Editing, Polishing''} [P93, P220], \textit{``Creative or Brainstorming, `culling the first main draft', draft editing, reviewing, finalizing''} [P113] \textit{``Outlining, Writing, Consulting''} [P158], \textit{``Schedule (content requirements), Reminders \& goals, Refresh''} [P135], \textit{``early drafting, end of drafting, reviewing''} [P224], etc.

In probing the ways our participants described these phases of writing we noted an overarching temporal ordering to these phases as well as a variety of activities that are associated with each phase.

\begin{quote}
    \textit{``
    Assistance for spellcheck/alignment  is much better suited to the end of \textbf{drafting}. Reminders to share or tag people are better suited for \textbf{reviewing}.''} [P224]\\
    \textit{``\textbf{Outlining} should help you structure your document before you begin.  \textbf{Writing} should check for spelling and grammar.  \textbf{Consulting} should make sure information is accurate.''} [P158]\\
    \textit{``The assistant could provide different services at \underline{different stages} \underline{of the doc}, such as during \textbf{editing} would provide with content suggestions; while \textbf{reviewing} it could provide with hints to include the right audience; on \textbf{publishing} it could help with Permission settings and privacy/compliance reviews''} [P50]
\end{quote}

A second round of qualitative coding of these responses resulted in identifying different activities that are associated with different phases of writing as perceived by our participants:
\begin{itemize}
    \item \emph{\textbf{Planning/Brainstorming Phase:}} Outlining/Diagramming [P36], section naming and setup [P46], help with structuring the document [P158]
    \item \emph{\textbf{Drafting Phase:}} Getting content or wording help [P44], structuring support to offload the mind [P63], support for sharing [P167]
    \item \emph{\textbf{Reviewing / Consulting Phase:}} notifications for formatting, grammar and spelling errors [P40], look of the document [P12], proofing [P75, P224], repetitive language usage [P167], highlighting unaddressed comments [P37], hints to include the right audience [P50, P133, P167], reminders to share or tag people [P224], fact checking [P123, P158], adding required references and sources [P133]
    \item \emph{\textbf{Finalizing / Publishing Phase:}} checking for broken paragraphs or tables before printing [P42], help with permission settings [P78], help with requesting required privacy and compliance reviews [P50]. 
\end{itemize}


\subsubsection{Modes of Writing}
Writing documents usually spans multiple sessions of activity that are distributed over document lifetimes. 
When a writer attends to a document (representing the workspace), to do a particular task, they bring with them a general understanding of the current stage of a document (more clearly visible in SAW scenarios) and a basic idea of what to look for or what activity to do (see also Gutwin and Greenberg's \cite{gutwin2002} descriptive framework of workspace awareness for real-time groupware).



Our participants referred to different goals they have when they attend to a document at different times, and that based on this goal they are seeking different types of support from the system.
A system can provide support that matches the intended activity: For example, hiding comments and change tracking annotations when the user is not in ``review mode'' or disabling editing when the user is in ``reading mode'' can help simplify the user's interaction with the writing environment and minimize accidental changes.

However, once we move to multi-author scenarios there is a need for a shared awareness of the document as the workspace, and the intent of a collaborator is impacted by this awareness.

\begin{myquote}{0.2in}
    \textit{``An ideal collaborative writing environment would allow users the choice of whether collaborators can view only or make edits to a document because the document should not be editable in certain situations.''} [P206]
\end{myquote}


\subsubsection{Awareness of Document's Stage}
So far, the three themes discussed above highlighted the importance of the detection of the current stage of a document or its corresponding writing phases to be able to provide writing support that is tailored to this stage. A distinct and interesting theme that was discovered through the qualitative coding of our survey responses was that the awareness of the stage of document's evolution is not only crucial for the system, i.e., an intelligent writing assistant, but is also crucial for collaborating authors who join a shared document at different stages of its lifetime. 
In single author scenarios an author who attends to a document at some point in time generally has a good understanding of the current stage of a document and what the next steps should be, whereas in collaborative writing scenarios developing and maintaining this workspace awareness is challenging \cite{gutwin2002}.
\citet{gutwin2002}, likewise, motivate the importance of workplace awareness for collaborative scenarios which can reduce effort, increase efficiency, and reduce errors for the activities of collaboration.
Assisting others with their local tasks is an integral part of collaboration and one that also benefits from workspace awareness. Awareness is useful because it helps people determine what assistance is \textit{required} and what is \textit{appropriate}. 



Some of our participants were proactive in suggesting ways that an intelligent system can detect the current stage of a document and share that information with other collaborators who attend to the document at different times:

\begin{myquote}{0.18in}
    \textit{``[The system can track] the number of people with the document opened for edit.  Tracking the "shares" of the document to know when it's sent for review or edit. the intelligent assistant that provided me the prompts to share it for review or editing would know [the stage of the document] from the selections I make.''} [P172]
\end{myquote}

\subsubsection{Summary of Findings}
Our survey data provided a range of insights regarding how authors perceive the stages of a document evolution and what types of writing assistance is expected at these stages of document authoring.
The qualitative analysis of survey responses indicated that the writers in our population perceive a document stage in three different, yet related ways: (1) temporal stages of document evolution; (2) phases of writing, and (3) modes of writing. 

\rvs{While phases of writing have been identified in past work, one thing that past work has struggled with is the link between phases of writing and document lifespan, i.e. the evolution of the document. Teevan et al. \cite{teevan2016} note that writing phases and writing workflow is ideosyncratic, and frequently interleaved; their solution is to break the writing process down into microtasks to support more effective collaboration. However, given that authors of multi-author documents (in our sample) perceive a link between document lifespan/document evolution and writing phases, one additional question that we can ask is whether or not different patterns of writing interactions can be identified over document lifespan such that differences in user behavior (e.g. commands used) could be associated with the stage of document evolution.} 
\rvs{To explore this question, in the next section we perform a log-based analysis that seeks to establish this link between the temporal stage of document evolution and differences in writer behavior.}


\rvs{\section{Characterizing Temporal Stages of Document Evolution}\label{QuantitativeStudy}}

\rvs{Ideally, one would be able to infer writing phases \textit{a priori} and through logs of interactions, but to the best of our knowledge there is no reliable way to directly measure authors' intended actions while interacting with a writing application.  However, given the perceived link in our survey responses between phases of writing and document lifespan,} in this section, we examine the temporal stages of document evolution through a different lens, by characterizing differences in user interaction logs of a commercial writing application. 
Our main goal with this quantitative analysis is to verify our qualitative findings at Web-scale and through a record of actual users' interactions as they are writing documents.

\rvs{In particular, we are interested in investigating three main research questions corresponding to the first three themes emerged from our qualitative data:
\begin{itemize}
    \item[$\textbf{[RQ1]}$] Do authors interact with the writing application differently at different stages of document lifetime? 
    \item[$\textbf{[RQ2]}$] Are there noticeable differences in the likelihood of observing certain activities at different points during the document lifetime?
    \item[$\textbf{[RQ3]}$] Do different authors attend to a shared document with different intentions as manifested in their first activities?
\end{itemize}
}

\subsection{Log Data Analysis}
\label{LogDataAnalysis}
We analyzed a sample of the anonymous interaction logs from users of \cmr{Microsoft Word} 
over a three-month period from February 1 to April 30, 2019. \rvs{We note that these interaction logs were available through an existing telemetry system. They were not specifically designed to understand user behavior but rather were designed to measure feature usage and performance for the corresponding writing application\footnote{
Users consented to log collection as a part of the user agreement}.} 

\rvs{To create a dataset that is suitable for investigating our main research questions specified in Section \ref{QuantitativeStudy}}, we collected a subset of document candidates in English from the US market that were created within a 2 week period from February 1 to February 15. For these documents we collected logs of user interactions with different functionalities available through the application UI.
\footnote{\cmr{While our analysis is based on the documents that are written in Microsoft Word, we note that this modern, widely used word processing application provides standard functionalities including copy-paste, formatting, commenting, change tracking, etc., which are available through other modern writing applications as well.}}
Our sample consists of approximately 16 million document candidates and approximately 16.6 million unique writers who accessed these documents.

\cmr{Note that the interaction logs are devoid of textual data and writer data.}
The only exception to this was a system generated ID assigned to an individual writer that served to uniquely map the set of interactions of that writer on a document. 
\rvs{This ID was not linked to any other identifying user information at any point in time, ensuring that the researchers had no information that could be traced back to an individual user.}
These interaction logs are highly granular and include approximately 2000 unique commands. For a subset of these documents \rvs{(96\% of initial documents)}, we generated numeric statistics on document content (e.g., \texttt{\#}words,  \texttt{\#}sentences,  \texttt{\#}charts, etc), logged as static snapshots at different time points.
Finally, for a much smaller subset of documents \rvs{(about 2\% of all docs)}, we have access to some statistics on typing traces that includes timing information on the duration of typing by different users with respect to a document\footnote{Since only a small subset of documents have typing statistics we only use this information when available.}. We do not have access to keystroke-level data or the actual characters that were typed \cite{birnholtz2013}.

\begin{figure}[htb]
\centering
\includegraphics[width=0.75\textwidth]{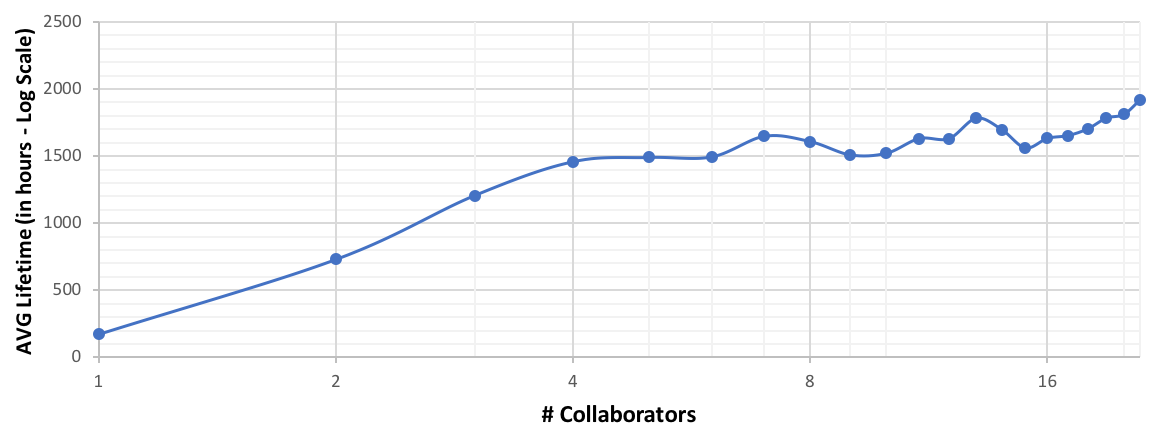}
\caption{Distribution of average document lifetime across different number of collaborators.}
\label{Lifetime_vs_numUsers}
\end{figure}
\subsubsection{Lifetime of Documents}
\rvs{Because a document evolves over its lifetime}, we 
defined the document lifetime as the time interval from its creation to the last user interaction with the document during our three-month window.  
We find that the majority of documents in our collection have a short lifetime (less than 24 hours - See Figure \ref{LifetimeDist} - gold/left).  In addition, there is a correlation between the number of collaborators involved in a document and the lifetime of the document. Figure \ref{Lifetime_vs_numUsers} shows that the lifetime of documents has a non-decreasing trend as the number of authors increases ($r = 0.78, n = 20, p < 0.0001$). 

\subsubsection{Pre-processing} 
\label{FinalDataSet}
We take two pre-processing steps to address limitations in our log data and prepare it for more robust analysis. 
Firstly, logged interactions are at a very fine-grained command level: we have over 2000 unique commands in our logs, each of which can be relatively sparse in its occurrence. 
To address this, we need to provide a mapping between low level user actions to higher level activities, which are interpretable and correspond to meaningful writing behavior.
\rvs{Initially, we looked to related past work \cite{birnholtz2013} to develop a taxonomy of commands; however, on inspection, we found these past taxonomies (e.g. insertion/deletion or major/minor edits) insufficient to fully describe the potential categories of commands in our dataset.}  To develop a richer set of initial categories of activities, a subset of all commands were reviewed by three annotators independently to identify the impact of applying those commands to the writing environment. The impact of a command was investigated based on widely accepted operations that change the content of a document (e.g., adding or editing) as well as commands that have no impact on the document content (e.g., viewing, navigating, communicating, etc).

\begin{figure}[htb!]
\centering
\includegraphics[width=0.75\textwidth]{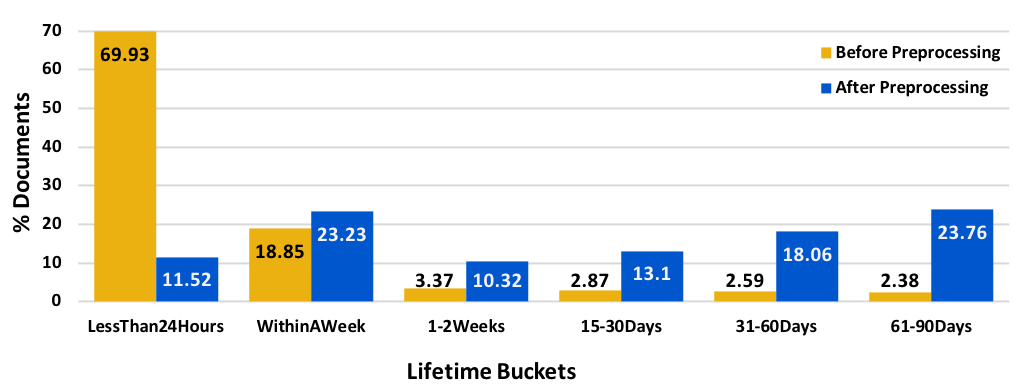}
\caption{Distributions of document lifetime before (gold - left) and after (blue - right) pre-processing} 
\label{LifetimeDist}
\end{figure}

\begin{figure}[htb]
\centering
\includegraphics[width=0.70\textwidth]{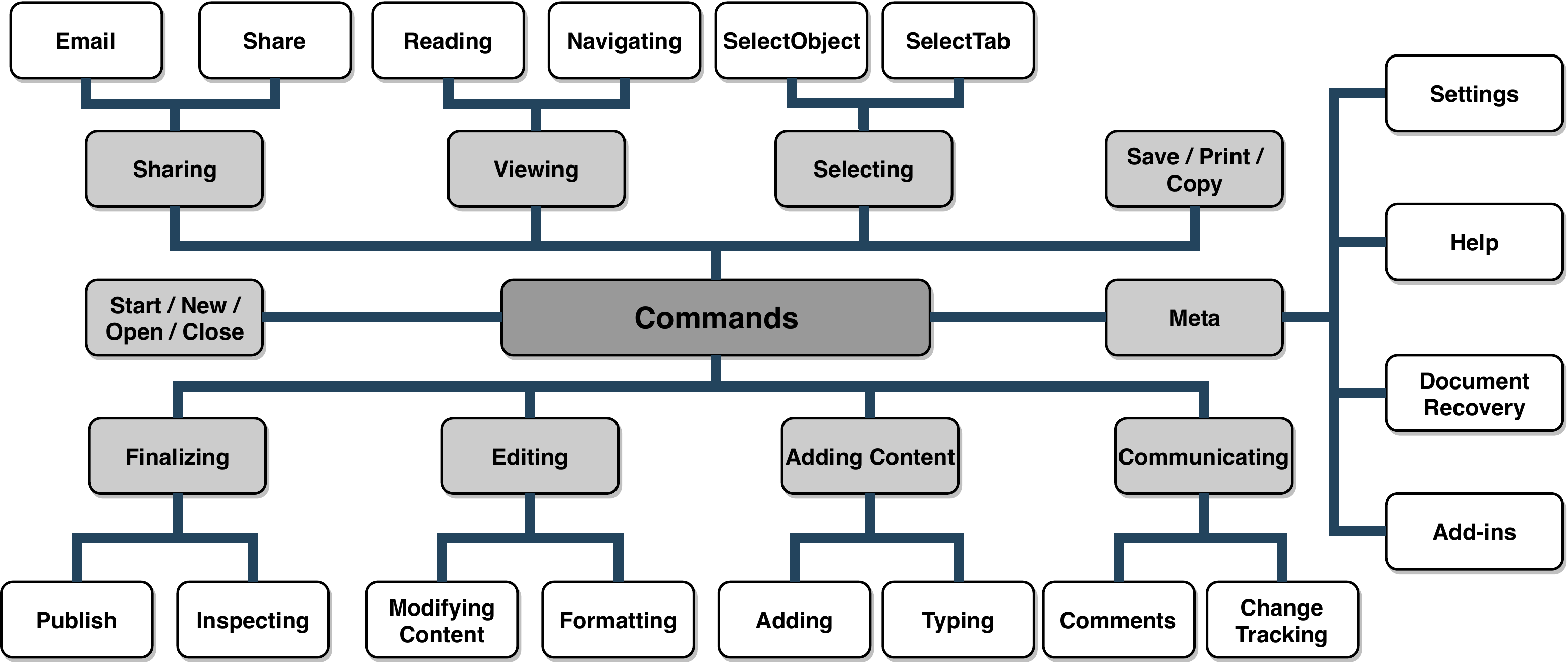}
\caption{Taxonomy of writing commands in our dataset. Note that only the top two levels are shown.}
\label{CategorizingCommands}
\end{figure}

The annotators then engaged in an iterative vetting and refining process \cmr{-- described by Campbell et al. as \emph{negotiated agreement}  \cite{CampbellCoding} --} 
to group commands based on how they impact the document and the writing environment. 
For example, all commands that result in adding some content (text, shapes, citations, etc.) to the document are grouped together as \textit{Adding}, while commands that are only used for navigating throughout the document (e.g., NextHeader, GoToPreviousPage, etc.) are grouped into the \textit{Navigation} category.
\rvs{Initial annotator agreement was moderate \cite{CampbellCoding}, at 67\%\footnote{\cmr{Note the use of percentages to describe inter-coder agreement, as per Campbell et al. \cite{CampbellCoding}.}} majority agreement between the three annotators.}
\rvs{\cmr{To enhance agreement through negotiation \cite{CampbellCoding}}, command categories were then revised iteratively for this subset of commands until all conflicts were resolved and the agreement was 100\%. This resulted in a finalized schema for grouping commands which was used by one of the authors to complete the groupings for the remaining commands in our logs while all authors would meet periodically to check for consistency.}
This analysis resulted in a set of \rvs{27} command categories. These categories were then classified into \rvs{10} higher level categories that more closely correspond to writing activities. \rvs{Figure \ref{CategorizingCommands}} shows a subset of the higher levels of this taxonomy. These high level groupings are sufficiently general and are supported by many modern writing tools.
\rvs{For more details and examples of the categories of commands in our dataset please see Appendix \ref{Appendix_commands}.}


The second pre-processing step is used to filter out all documents with a single author throughout their lifetime. While writers in individually authored documents do progress through a similar yet simplified subset of writing activities (as characterized by Lowry et al. \cite{lowry2004}), we are focused on characterizing stages of document evolution where multiple authors write collaboratively.  
Thus we limit most of our analyses to documents that log interactions from 2 to 10 users\footnote{Our survey confirms (see Section~\ref{QualitativeFindings}) that 98\% of collaboratively authored documents have no more than 10 users.}. Note that our methodology is, in general, agnostic to the number of users, and
throughout the rest of the paper we will show some results that equally apply to individually authored documents.

Finally, in order to reduce noise in the dataset, we eliminate documents that have a lifetime shorter than 1 hour.
Our final dataset comprises approximately 170k documents where the majority has 2 or 3 collaborators involved (78\% and 12\% respectively). The lifetime of documents in this final set are higher and more evenly distributed overall (Figure \ref{LifetimeDist} - blue/right).

\subsubsection{Analysis.} In our quantitative analysis we focus on quantifying how authors interact with the writing application over temporal stages of a document lifetime. \rvs{In order to facilitate the analysis of interactions with the writing application over time, we divide the lifetime of a document into equally-sized segments that correspond to 10-percent buckets of the document's lifetime. We note that a linear segmentation of the document's lifetime does not imply writing phases can be discretized as fixed-size, non-overlapping buckets. Rather, the goal of our quantitative analysis is to understand changes in the patterns of interactions with the application over time. Past research coupled with our qualitative findings suggest that authors perceive a temporal ordering to certain phases of writing (e.g. planning, drafting, copy-editing) and associate some writing activities to early stages of document lifetime and other activities to later stages of document evolution.} 
Overall we found that writers distribute their writing activities
differently over stages of document authoring both in
terms of magnitude as well as types of these activities. 
\cmr{Further, through an analysis of the distribution of authors' first activities we highlight differences in the authors' modes of writing.} The following \cmr{three} subsections describe these differences in detail.

\subsection{Magnitude of Writing Activities Over Time} \label{Data:Catplots}
We first characterize a document's temporal stages by tracking the magnitude of author interactions with the writing application over a document's lifetime, hereinafter referred to as Contributions Across Time (CAT). To this end, we segment the lifetime of each document into 10 equally sized buckets and look at the magnitude of interactions in each of these segments.

\begin{figure}[bht]
  \centering
  \subcaptionbox{Single vs Multi Author Docs\label{fig:SAD_MAD}}{
    \includegraphics[width=0.47\textwidth]{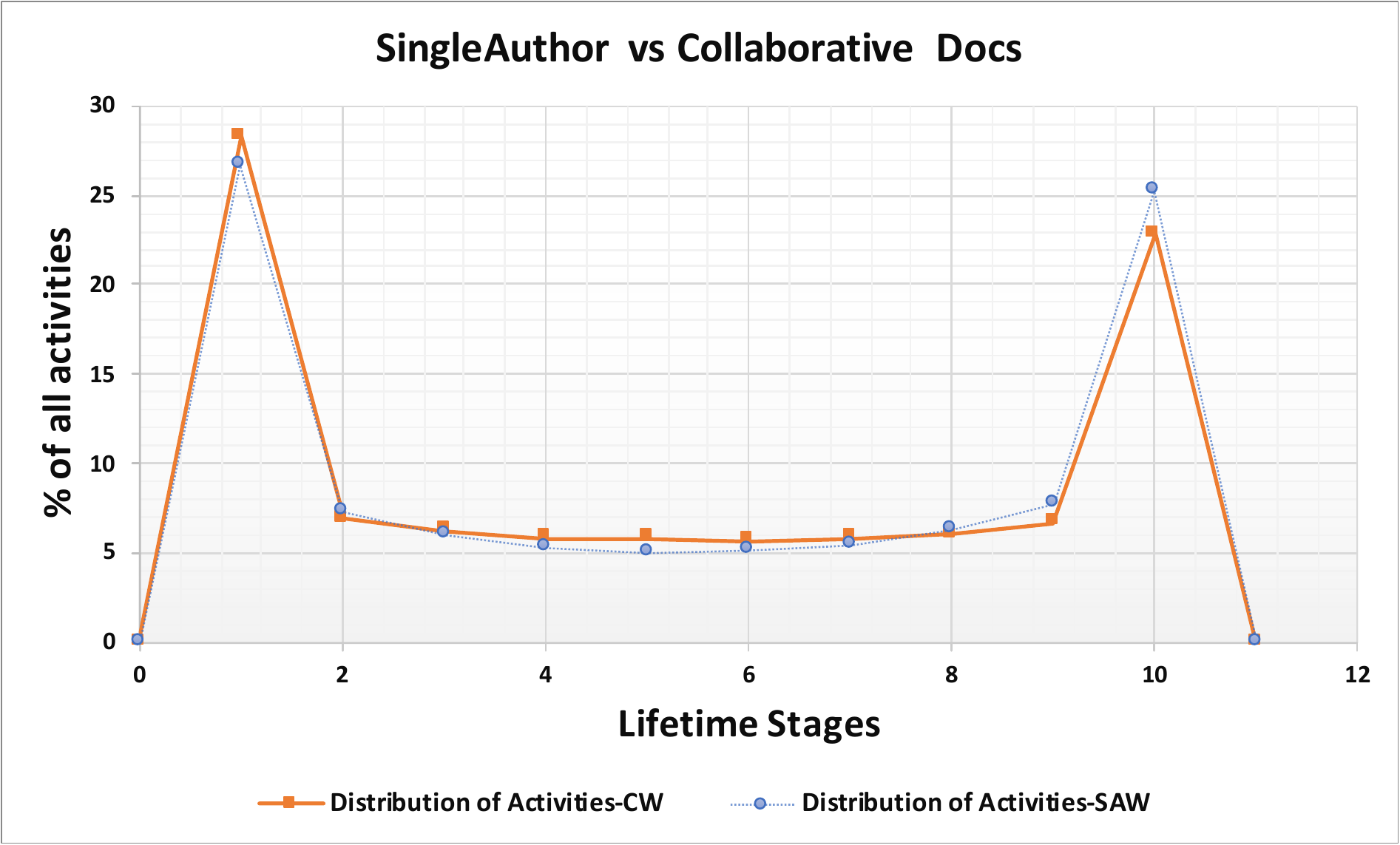}
  }
  \subcaptionbox{Different Lifetimes\label{fig:MAD_Lifetimes}}{
    \includegraphics[width=0.47\textwidth]{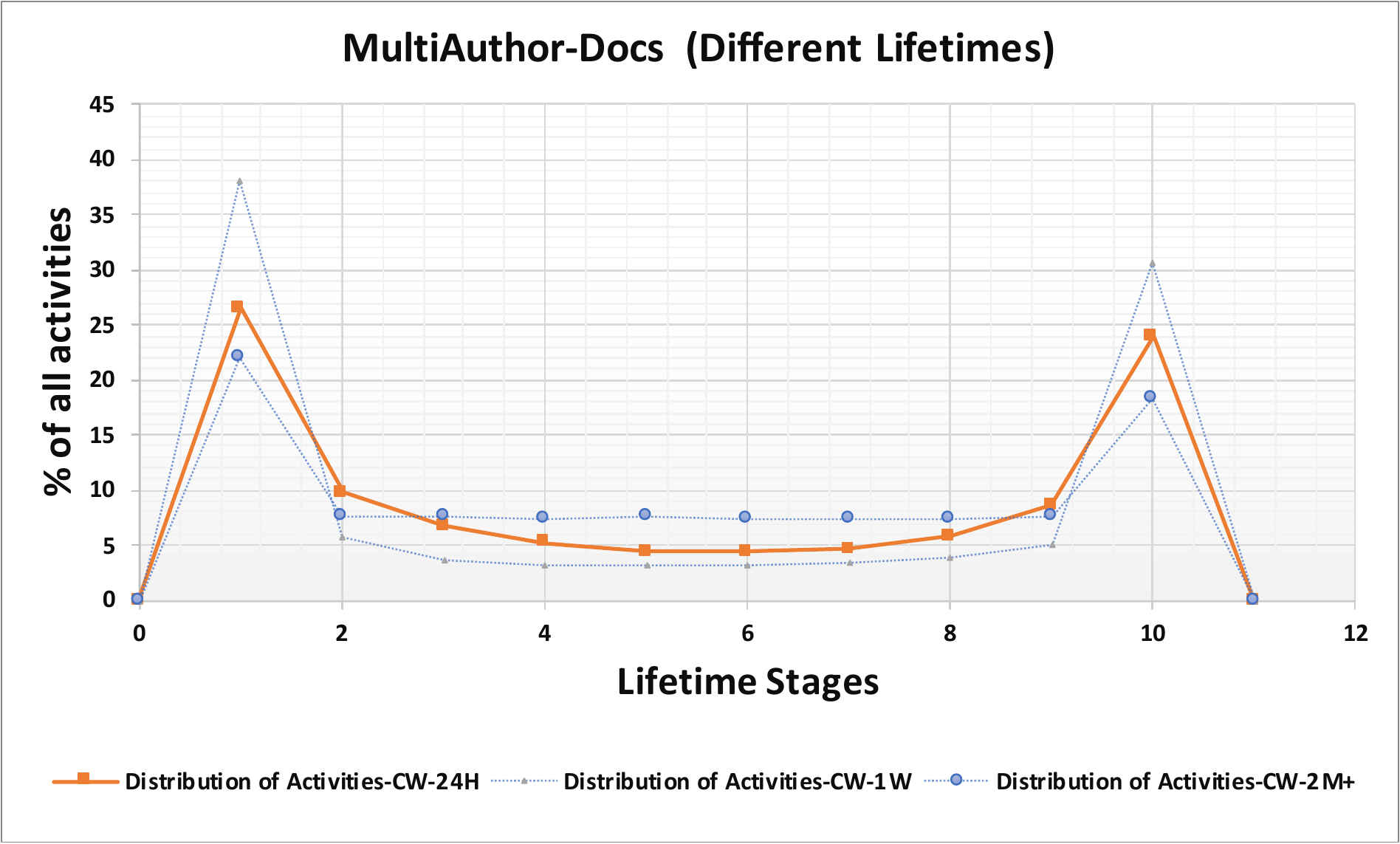}
  }
  \subcaptionbox{Different \texttt{\#} Collaborators\label{fig:MAD_numCollaborators}}{
    \includegraphics[width=0.47\textwidth]{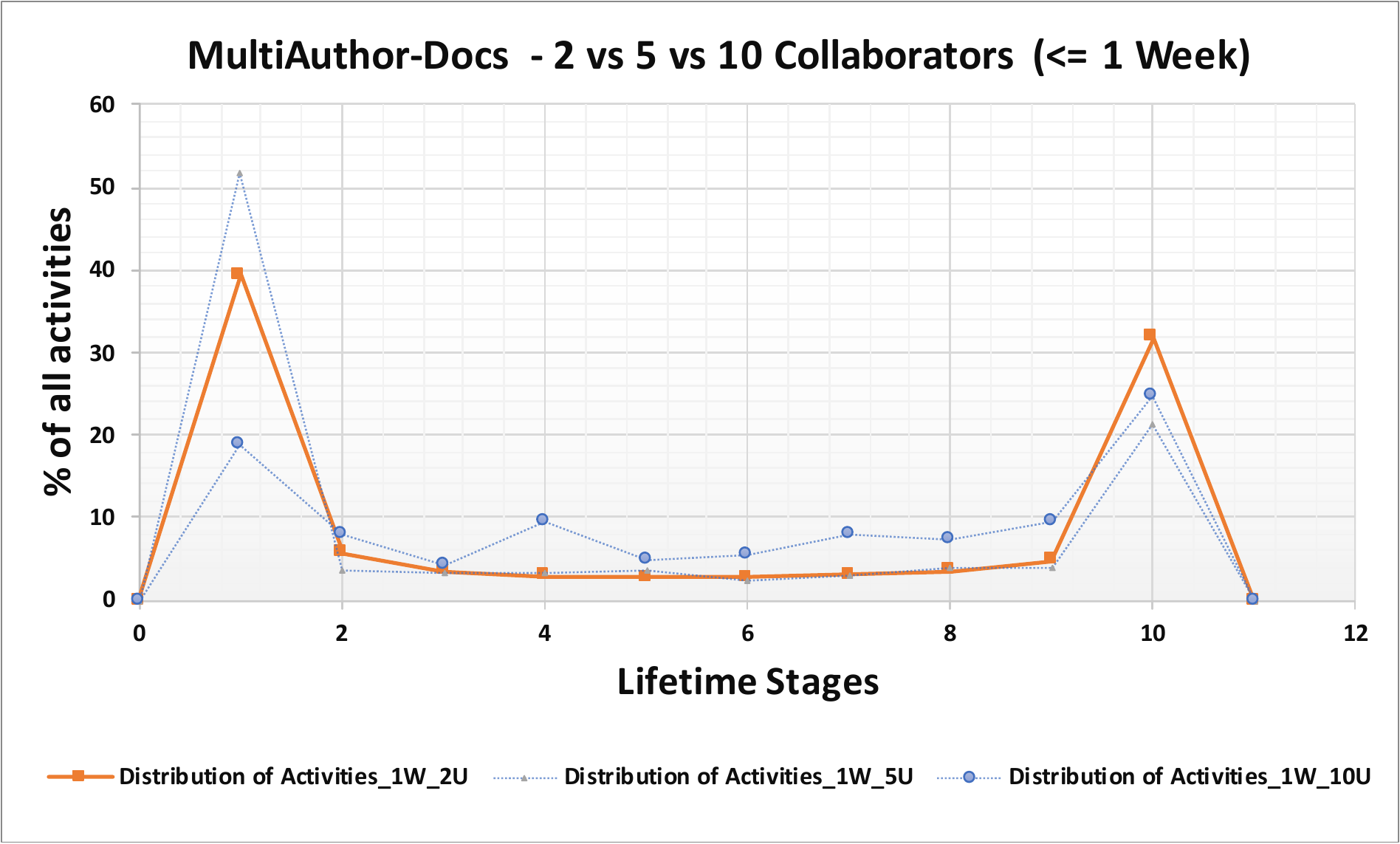}
  }
  \caption{Distribution of frequency of writing activities over stages of document evolution (a) single vs multi author documents, (b) multi author documents with different lifetimes, and (c) multi author documents with different \texttt{\#}collaborators.}
  \label{fig:Catplots}
\end{figure}

Figure \ref{fig:Catplots} shows these CAT distributions for different types of documents.
These activities are normalized by the total frequency of activities at the end of the lifetime of the documents. \footnote{The x-axis corresponds to the 10 buckets over the lifespan of the document plus two additional buckets (pre and post) for before the document is created (0-operations) and after document editing ends (0-operations), yielding 12 points in Figure-4.}  We can see an interesting pattern where there is a high level of activities at the very early and very late stages of document writing, while the middle stages show a much lower level of interaction with the writing application overall.

Interestingly, even when we move away from single-author documents to multi-author documents with different numbers of collaborators (See Figure \ref{fig:SAD_MAD}), the overall pattern does not change. We also experimented with documents of different lifetimes (i.e., short [between 1 and 24 hours]; medium [within a week]; long [more than 2 months] - Figure \ref{fig:MAD_Lifetimes}), as well as different number of collaborators with the same lifetime (Figure \ref{fig:MAD_numCollaborators}) and noticed the same pattern persisted throughout.  

The progression of these writing activities over the document lifetime is interesting: there is a notable difference at early and later stages of document authoring that could be attributed to the task management aspects of document writing and how authors distribute the workload over the course of the writing process. Another implication is in predicting the lifetime of documents by observing the changes in the rate of interactions between the users of a writing application and the writing environment. We revisit the problem of predicting the document's lifetime in Section \ref{predictionSection}. 

Please note that these CAT distributions can only be observed at the end of the document lifetime where we can compute the percentage of activities at different stages of the document evolution. That is, throughout the document evolution we can only capture the rate at which the interaction level is changing and compare it against previous snapshots of the document. 
\subsection{Types of Activities in Early vs Late Stages}
\label{Logs_TypesOfActivities}
\begin{figure}[htb!]
\centering
\includegraphics[width=0.95\textwidth]{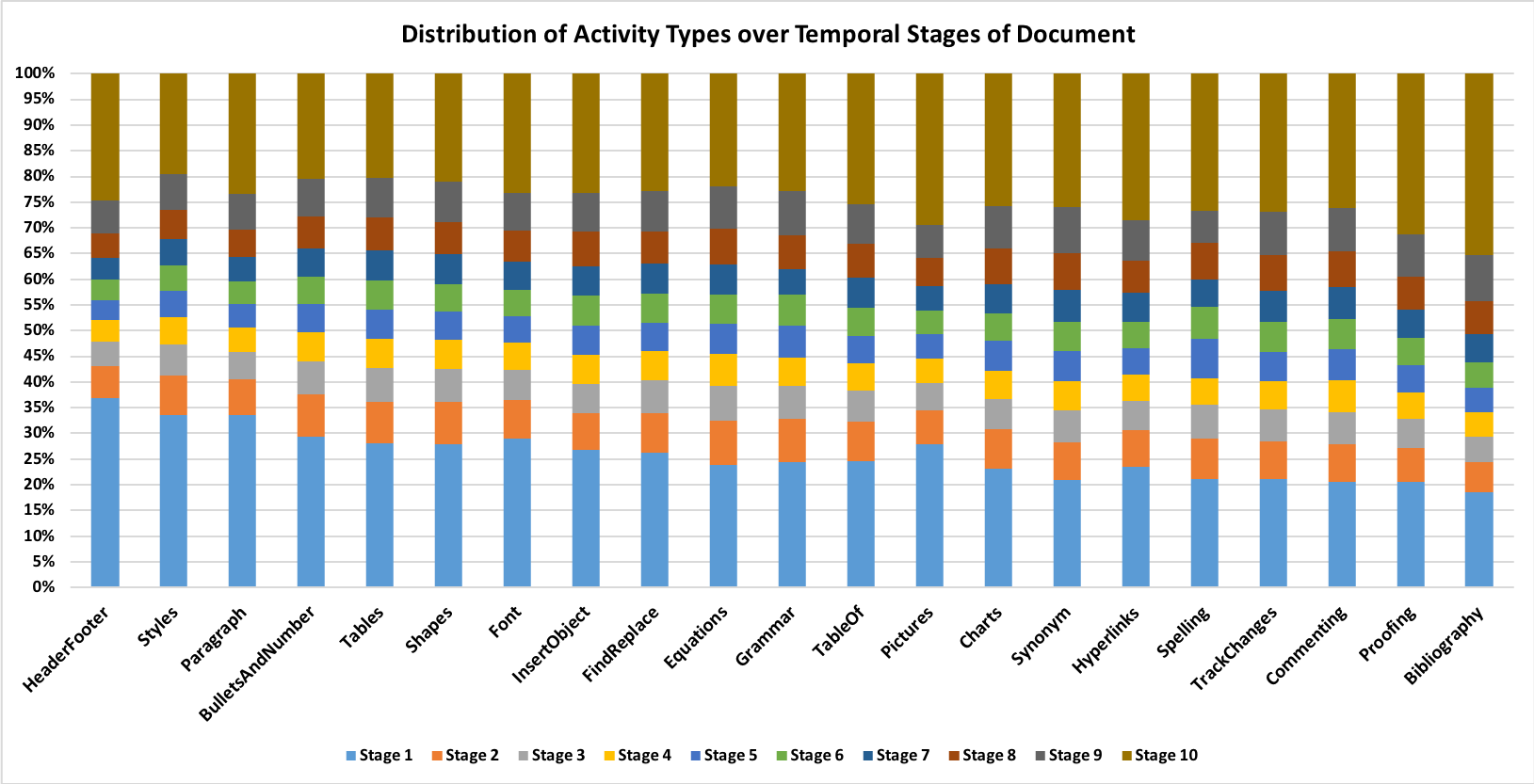}
\caption{Distribution of types of writing activities over stages of document evolution.}
\label{ActivityTypeDist}
\end{figure}

The next research question we investigated was whether our qualitative observations regarding the types of activities that writers perform in earlier versus later stages can be observed in our large scale dataset of user interactions.
Recall that based on our survey responses, outlining, structuring the document and drafting were among the early activities, while proofing, collaborative activities, referencing were identified as later stage activities.

Reusing our lifetime segmentation of authors' contributions we looked more closely at the distribution of different types of writing activities across these 10 segments.
A subset of activities that correspond to the ones mentioned by our survey participants was used for this analysis. \rvs{Please note that this set of activities correspond directly to the ways that different functionalities are grouped by the writing application UI. These groupings provide a different categorization of users commands that is complementary to the categories derived iteratively and exhaustively from our commands as discussed in Section \ref{FinalDataSet}.}
Figure \ref{ActivityTypeDist} demonstrates the distribution of these activities over these 10 stages of document evolution. These activities are ordered from left to right based on the difference between their shares in Stages 1 and 10 to reflect how early they are likely to be observed during the document evolution. 

The distribution of each of these activities over the 10 stages of document lifetime is normalized by their overall frequency. Note that for each of the activity types, stages 1 and 10 are the most prominent stages in which they occur, which reflects the results from the CAT plots.
However, when comparing the likelihood of a writing activity between the first and the last stage we can see variations across activities. 
For example, activities such as interacting with HeaderFooter, Styles, Paragraph, and BulletsAndNumbers are more likely to be observed during the earliest stage of document authoring than the latest stages (e.g., 37\% vs 24\% for HeaderFooter; 34\% vs 23\% for styles and paragraphs). These activities correspond to setting up the overall structure of the document, outlining and early stages of drafting which involves creating and modifying lists and working them into paragraphs. A variety of advanced objects including Tables, Shapes, Equations and Charts are added and manipulated throughout the middle stages, with Tables and Shapes preceding Equations and Charts. 

Among the later stage activities we can see improving the writing style (e.g. finding Synonyms) is more likely to be a later stage activity (26\% vs 20\%). Collaborative activities such as TrackChanges and Commenting are also likely to happen towards the end of a document's lifetime.
Interestingly interacting with Proofing and Spelling functionalities are 11\% and 6\% more likely to happen at the latest stage, while Grammar related interactions seem to be middle stage activities. Finally, once we examine specific proofing commands we see some evidence of \textit{copy-editing mode} that seems to be more likely in later stages. For example, 28\% of all occurrences of NextMisspelling command, representing an author sequentially navigating through spelling suggestions by the writing tool, was observed at the latest stage (vs 22\% during the first stage).  
\subsection{\cmr{Distribution of First Activities}}
\label{Logs_FirstActivities}
For each document in our dataset we also looked at how activities differ between authors that start working on the document at an earlier or later stage. For this purpose, we ordered the authors of a shared document by the time of their first activity in the document and we collected the type of their first activity as well as the set of all their activities.
Contrasting the distribution of activities by the first and second authors of a shared document we do not see a high variation. Overall, Formatting, Adding and Modifying Content accounts for 22\%, 15\% and 15\% for both first and second authors. The main two distinctions are for Typing and Initiating Communication (e.g., adding a new comment) that accounts for 5\% and 0.7\% of second author's overall activities but only to 2\% and 0.4\% of first author's overall activities.  

Once we shift our focus to the first activities of the first and second authors at the time they first attend to a shared document, we start noticing differences in their intent and modes of writing. Our log based analysis indicated that the second authors are $7\times$ more likely to start their interaction with typing and $2.5\times$ more likely to start with adding non textual content (e.g., advanced objects) compared with the first authors (14\% vs 2\% and 10\% vs 4\% respectively). 
A similar trend is observed with second authors being two times more likely to start with editing activities (i.e., modifying content and formatting - 10\% vs 5\%). 
Finally, not only do second authors initiate communication more than the first authors throughout the document lifetime, they are also three times more likely to attend to a shared document for the first time with a communication intent (i.e., commenting and change tracking activities).   


\subsection{Summary of Findings}
From our qualitative data, we find that authors have a perception of stages or phases in the writing process and that these stages represent an ordering in the creation of a document -- from ideation through creation to refinement. Our quantitative data reflects this partially ordered approach to document creation through changes in both the magnitude of and types of interactions that occur over a document's lifetime.
Further, authors in our qualitative study welcome the idea of stage-appropriate writing assistance. They highlight varied goals and activities to which assistance can be tailored at different stages. Alongside the potential for stage-specific assistance, previous work has noted the importance of stage-appropriate behavior during collaborative tasks \cite{gutwin2002}, and therefore stage-aware tools have the potential to foster improved collaboration. 

In summary, implications of the synthesis of our qualitative and quantitative data is that \begin{enumerate*} \item there exists a perceived benefit to stage-specific writing assistance, \item we observe different activities at different stages of the document lifespan, and \item stage awareness mechanisms can communicate to collaborators the current state of the document, which can aid in on-boarding of new collaborators and in appropriate behaviors during collaboration \cite{gutwin2002} \end{enumerate*}. However, to effectively design interfaces to support stage-specific writing tasks and to communicate stages, one open question remains: Can we predict \rvs{the stage of document completion} given observation of user actions? To explore this last research question, we present a study in the next section assessing the feasibility of
stage-identification. 

\section{Predicting the Stage of a Document}
\label{predictionSection}
Our qualitative analysis has shown that users would appreciate a more situated writing assistant that takes the stage of a document into account. To this end, a system that could predict the stage of a document automatically would be of high value. From our log analysis we also know that document evolution is reflected in the signals we find in user interaction logs. In this section, we conduct a preliminary investigation into how useful log signals can be for predicting a document's stage. Specifically, given a telemetric snapshot of a document -- at a uniformly and randomly sampled point in its lifetime -- we attempt to predict the quartile of the document's lifetime to which the snapshot belongs. We chose to represent the document lifetime in only four stages for this preliminary investigation to simplify the task. Note that despite this simplification, this is a non-trivial prediction task, since the lifetimes of individual documents vary greatly in our data, as do the magnitudes and distributions of their logs.
\cmr{We note that being able to effectively identify boundaries between phases of writing or detect transitions between these stages would be a more interesting goal. However, to the best of our knowledge the most successful attempts at inferring phases of writing (e.g. planning versus drafting versus reviewing) was achieved by analyzing keystroke logs and used the time intervals between different keystrokes to infer changes in the users' cognitive load and correlate them with phases of writing they are transitioning through (e.g. \cite{conijn2019}). Given issues of data privacy, our goal was to leverage interaction logs that are devoid of any text and keystroke information.}

We sample the longitudinal logs of 60k documents from our collection of 170k documents (see Section~\ref{FinalDataSet}). 
We use 50k of these documents for training and reserve the rest for testing. For each document, we sample five distinct \emph{snapshots}, each of which is an aggregate of telemetry signals starting with the document's creation and ending at a uniformly and randomly sampled point \emph{t} in the document's  lifetime. We represent \emph{t} as a float number in the range [0:1] where 0 corresponds to the beginning of the lifetime and 1 corresponds to the end of the lifetime. Each document's lifetime is divided into 4 equally sized buckets with the snapshot boundaries 0, 0.25, 0.75, 1.0. Given this, any snapshot can be labeled by its corresponding quartile by mapping the float number to the quarter bucket it belongs to.  
Given this data we train and evaluate four one-versus-all boosted decision forest classifiers\footnote{We experimented with several types of classifiers, including logistic regression, SVM, and multi-layer perceptron, and found boosted decision forest to perform the best.} \cite{friedman2000} -- one each for the four lifetime quartile labels in our dataset.
The model is trained with telemetry features from the lifespan of the document up to the timestamp the snapshot is sampled from. In other words, the model has no access to information beyond this point or any features from the entire lifespan.

\rvs{Our model incorporates features from the user interaction logs that can broadly be grouped into the following five classes:
\begin{enumerate*}
    \item commands
    \item command categories
    \item high-level command categories
    \item advanced functionalities
    \item content statistics
\end{enumerate*}
(see Sections~\ref{LogDataAnalysis} and \ref{Logs_TypesOfActivities} as well as Appendix \ref{Appendix_features} for more detail).} The first four feature classes are normalized -- that is, individual feature values in each class represent the percentage of contribution of a feature within its class, at a given snapshot of a document. For example, the command category \emph{Navigation} will have as its value the percentage of navigation commands amongst all command categories observed in the snapshot. We supplement these signals with two sets of features 
designed to reflect the balance of collaboration in a document. Our qualitative data suggests that not all stages of document authoring are collaborative and therefore these features can provide some signal in predicting the stage of a document. 
These collaborative features measure the contribution of different collaborators to a shared document across 4 dimensions: adding content, editing, communicating, and finalizing (see Figure \ref{CategorizingCommands}). 


The first metric, Contribution Balance Score (CBS) broadly considers any signal logged to one of the 4 categories as \emph{content contribution}, and computes the share of \emph{content contributing} activity performed by each collaborator of a document.
The second metric, Role Balance Score (RBS), separates contributions into one of the 4 dimensions of activity, in order to determine whether collaborators took on dedicated roles (e.g., Writer, Editor, etc.) over the course of document authoring. 
We adapted the method used in \cite{pace2018} to measure the contribution of different collaborators using their share of basic text-based operations to our interaction logs and computed the CBS and RBS features as follows:

\begin{equation} \label{eq:cbs}
    CBS(d_i) = \displaystyle \frac{1}{log(u)} \sum_{j=1}^{u} CC(ij) \cdot log \left( \frac{1}{CC(ij)} \right)
\end{equation}
\begin{equation} \label{eq:rbs}
    RBS(d_i, a_k) = \displaystyle \frac{1}{log(u)} \sum_{j=1}^{u} AC(ijk) \cdot log \left( \frac{1}{AC(ijk)} \right)
\end{equation}

where $u$ is the number of authors observed in a document snapshot, $CC$ the number of general content contribution signals, and $AC$ the number of specific activity contribution signals to one of the 4 dimensions $a_k \in \{\text{Adding, Editing, Communicating, Finalizing}\}$.

We compare our model against a baseline that uses the sole feature of elapsed time since document creation 
Note that this signal provides a strong baseline, especially for the last quartile of a document's lifetime. 
In particular, given that our data longitudinally tracks documents over a maximum period of three months, if the elapsed time exceeds, e.g., two months, it is highly likely that the snapshot under consideration belongs to the 4th quartile label.

\begin{figure}[htb!]
\centering
\includegraphics[width=0.65\textwidth]{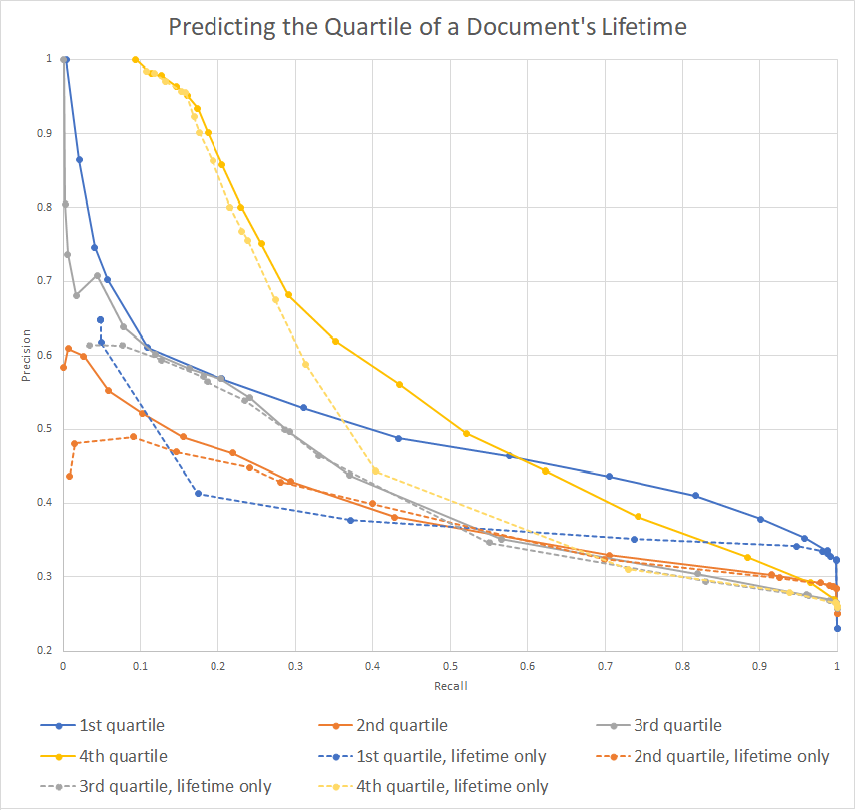}
\caption{Precision and recall curves for predicting the quartile of a document's lifetime. Curves are shown per model, per quartile. Dashed lines = time elapsed baseline, Solid line = learned model.} 
\label{PredictionPR}
\end{figure}

Comparing the baseline against a model trained on all features, we note that the baseline achieves a macro-averaged accuracy of 41.77\% and the model with all features outperforms this score with 44.80\%. Differences are significant at p < 0.01 (Approximate Randomization Test \cite{edgington1969approximate}).
Figure \ref{PredictionPR} shows the precision-recall characteristics of our models for each of the four predicted stages. Precision and recall are computed per predicted  quartile to illustrate differences in the model performance across quartiles. The chart in Figure \ref{PredictionPR} contains a total of 8 precision recall curves, for the combination of two settings (using lifetime features only versus using all features) and four labels (1st, 2nd, 3rd, and 4th quartile). 
Overall, the results demonstrate that user interaction logs do contribute positively towards the prediction of a document's temporal stage. Interestingly, the most significant gains are observed for the first and last quartile labels in documents' lifetimes. These seem to correspond to the CAT distributions 
that we observe for magnitudes of writing activities over a document's life span (see Section~\ref{Data:Catplots}, Figure \ref{fig:Catplots}). We hypothesize that when user interaction logs are recorded in volume they can provide a valuable additional signal to predict the temporal stage of documents.

We also perform manual feature analysis to identify the most predictive features across classifiers. Not surprisingly, elapsed lifetime is at the top of the list, 
followed by the CBS and RBS collaboration features and then a mix of the other signals, from individual commands to activities and activity categories. 

In summary, our first attempt at modeling a document's temporal stage has shown promising results with signal from user interaction logs. This is nevertheless, a preliminary foray into a task that could almost certainly benefit from more complex modeling approaches. For example, a model capable of capturing sequential dependencies (such as an LSTM) will likely be better able to leverage the temporal changes in a document's evolution than our current approach, which aggregates the document's history into a snapshot. 
We leave these efforts to future work.
\section{Discussion and Design Implications}
Previous work \cite{freedman1987, fitzgerald92, flower1984} on supporting authors who write documents individually or with others, have characterized the process of writing as a series of stages  -- planning, drafting, reviewing and finalizing -- and that the decomposition of these stages can be leveraged by the authors as a means of scaffolding their work \cite{graham2012}, reducing their cognitive load \cite{kellogg1988, greer2016}, and supporting more effective coordination and collaboration \cite{teevan2016}.
While techniques to support collaborative writing (e.g., microtasking \cite{teevan2016}, effective communication and coordination mechanisms \cite{galegher1994}, or automated spelling and grammar support in modern writing tools such as Google Docs or Microsoft Word) appear promising, we would argue that any technique that seeks to leverage stages of writing requires, \textit{a priori}, an understanding of whether authors perceive stages in writing, of whether those stages generalize across a broad range of writing tasks, and of whether or not the stages of writing can be automatically inferred.

In our work, we specifically target this notion of staged document writing.  The three previous sections provide evidence that authors perceive stages and associate specific activities to them, that the magnitude of writing activities in both collaborative and individual writing follows a similar pattern in these stages, and that these stages can be inferred from privacy-preserving data logs of document interaction.  Together, these results provide support for the feasibility of stage-aware writing assistance.
More specifically, our qualitative findings provide a range of insights about different characteristics of temporal stages of document authoring: Earlier stages of writing were described as more focused and less structured, suggesting the necessity of minimizing system interruption and affording more control to the user; Later stages, on the other hand were perceived as more mature and more stable, where the system can be more proactively involved in providing local and global feedback and suggestions to improve the overall document quality. 

One significant and interesting observation from our log analysis was the similarity in ways that authors in both individually authored and multi-authored documents distribute their work load across stages of document evolution (see \cmr{Section \ref{Data:Catplots}}).  While initially unexpected, this similarity is important in the design of document writing assistants because, if these intelligent assistants are destined to guide author behavior through a determination of stages, then regardless of whether one is writing alone or collaborating, the overall goal of any intelligent writing assistant would be to support the writer(s), making them a default integration into digital writing platforms.  
In hindsight it may be unsurprising that individual and group document editing share some similar aspects: some past views \cite{rohman1965} of the stages of document creation have specified stages of pre-writing, writing, and re-writing, i.e., that individual authorship is a collaboration with oneself through the mechanisms of communication and distributed cognition \cite{zamel1982}. Another important implication of this observation is in the ability to infer the stage of a document evolution by simply tracking the rates of interactions between the authors and the writing tool.

In collaborative scenarios, our analysis of distribution of first activities (\cmr{Section \ref{Logs_FirstActivities}}) provided quantitative evidence for our qualitative characterization of modes of writing and indicated that first and second authors of a shared document enter this joint workspace with noticeably different intents and modes of writing. 
For example, we find that second authors of a shared document are 3X more likely to appear on the shared space with communication intent and 7X more likely to start their collaboration with typing new content. This finding has implications for designing more customized toolbars for the writing applications that can match the intended mode of writing (e.g. reviewing versus reading mode), which helps simplify the user's interaction with the tool and minimizes accidental changes.

Every study comes with different limitations. In our work, we also acknowledge at least two potential confounds: (1) our survey population is limited to information workers in a large technological company. We made an attempt to mitigate this limitation by recruiting across many different departments with a diverse set of job roles and demographics; our log analysis seeks to address this limitation by considering a much broader population of authors. 
(2) the prediction accuracy of our trained model is currently at an acceptable level, but it can be improved by leveraging candidate sequence learning methods (e.g. LSTM) which can capture sequential dependencies we observed in our data (i.e. CAT distributions). We still argue that our preliminary attempt towards inferring the stage of a document can serve as a promising starting point for designing stage-aware writing assistants.

While our work on linking the stages of document evolution to writing activities is important to the design of digital collaborative document authoring assistants, it is clear that much work remains to be done to make these tools a reality. For example, while we showed that the current stage of a document under development can be inferred and utilized to provide timely writing assistance that is tailored to this stage, how we realize tools to support stage aware writing assistance remains an open problem.  \rvs{For example, should tools automatically infer writing stage and provide varied assistance based on this inference?  Or should users pre-specify writing stage?  Intelligent adaptation of an interface is something that researchers have extensively probed \cite{Bunt2007, horvitz1999}, and the reluctance to automatically trust adaptive tools \cite{Bunt2007} was echoed in our initial survey, where over one third of participants were reluctant to have systems infer writing stage.  How to incorporate tool adaptivity into any interface that leverages inference to improve productivity remains an on-going area of research.  On the other hand, echoing past work in collaborative systems \cite{gutwin2002} and in collaborative writing \cite{teevan2016}, appropriate writing etiquette is also an important part of effective collaboration.  Therefore, another issue to probe with future work is that awareness of document evolution does not, necessarily, need to adapt the interface to different phases of writing, but can instead be leveraged to foster more effective collaboration through enhanced task awareness \cite{wang2019}.  In this way, instead of enforcing modified behavior on the writer, these tools can instead, serve to promote improved context awareness for collaborators in computer mediated collaborative writing activities.}

\section{Conclusion} \label{sec:concl}



In this paper, we have presented the first large-scale study of the stages of writing, specifically as they relate to the temporal evolution of documents. Using a mixed-methods approach that includes qualitative, quantitative, and predictive analyses, this paper presents a comprehensive picture of how authors perceive stages of document authoring, how these perceptions correlate with actual user interaction logs, and how these logs can, in turn, be used to predict the temporal stage of a document. More specifically, a survey of 183 participants concluded that writers do have mental models of definite stages while authoring documents; moreover, they associate different goals and types of activities with each of these stages. We support the findings of this qualitative survey with a quantitative analysis of writers' longitudinal user interaction logs on a popular Web-scale document authoring platform. These logs are notable in two important ways. First, they cover a set of several million documents -- thereby surpassing the scale of any previous study of document evolution by many orders of magnitude. Secondly, these logs contain no actual text -- thereby overcoming some of the assumptions about document topic or quality embedded in previous approaches. Our findings confirm that writers' activities are distributed differently through different stages in a document's evolution, both in magnitude and in the types of activities.

Finally, as a first step towards enabling stage-aware intelligent writing assistance, we conduct a predictive experiment to model the temporal stage of a document by featurizing the user interaction logs. The results of our experiments suggest that these logs provide 
predictive power over a strong baseline. These findings show promise for future work on intelligent writing assistance.
\vspace{-200mm}



\begin{acks}
We would like to thank Shamsi Iqbal, Justin Cranshaw, Mark Encarnaci\'on, Michael Bentley, and Tomasz Religa for their valuable input throughout this work. We also thank the reviewers for their feedback; and finally, all of our study participants for their help in making this research possible.
\end{acks}

\pagebreak
\bibliographystyle{ACM-Reference-Format}
\bibliography{writing_research}

\newpage
\appendix

\section{Description of Main Command Categories}
\label{Appendix_commands}
\renewcommand\tabularxcolumn[1]{m{#1}}
\begin{tabularx}{\textwidth}{m{3cm}|X}

\caption{Description of Commands Categories and Sample Commands per Category.}\\
\hline
\multicolumn{1}{c}{\textbf{Name}} & \multicolumn{1}{c}{\textbf{Description}}\\
\hline\endfirsthead
\endhead
\multicolumn{2}{r}{\itshape continues on next page}\\
\endfoot
\endlastfoot
    \rowcolor[gray]{.9}Adding Content & Commands in this category add to the content of a document by either inserting an object or an element to an object (e.g. Tables, Shapes, etc) or text (e.g. by pasting or typing). \\
    \hline
    SubCategories & Sample Commands \\
    \hline
    \multirow{5}{*}{Adding}
        & InsertCitation \\
        & InsertBibliography \\
        & InsertFile \\
        & Paste; PasteTextOnly; $\cdots$ \\
        & TableInsert \\
    \hline
    Typing & Typing is captured as system's events and we add these events to the list of commands that are logged based on the interaction with the tools UI to incorporate the periods of typing into our dataset \\

    \hline
    \rowcolor[gray]{.9}Editing & Commands in this category result in changes either to the content or the formatting / styling characteristics of existing content in the document. \\
    \hline
    SubCategories & Sample Commands \\
    \hline
    \multirow{5}{*}{ModifyingContent}
    & AutoCorrect \\
    & ChartDelete \\
    & Replace \\
    & EditCitation \\
    & JoinList \\
    \hline
    \multirow{5}{*}{Formatting}
    & listPosition \\
    & SetTransparentColor \\
    & Recolor \\
    & BibCitationToText \\
    & Bold \\
    \hline
    \rowcolor[gray]{0.9}Viewing & Commands in this category do not result in any changes in the content or formatting of the document. These commands are tracking users navigational or viewing interaction with the writing environment. \\
    \hline
    SubCategories & Sample Commands \\
    \hline
    
    \multirow{4}{*}{Navigating}
    & NextField / PreviousField \\
    & NextHeader / PreviousHeader \\
    & GoToHeader / GoToFooter \\
    & NextMisspelling \\
    \hline
    \multirow{4}{*}{Reading}
    & ViewReadingMode \\
    & ReadingModePageview \\
    & ExpandAllHeadings \\
    & ToggleReadingMode \\
    \hline
    \multirow{5}{*}{Viewing}
    & BibliographyFilterLanguages \\
    & DocInfoFileName \\
    & ShowDocumentText \\
    & ViewHeader \\
    & Find / FindNext \\
    \hline
    
    \rowcolor[gray]{0.9}Selecting & Similar to the Viewing category, commands in this category do not result in any changes in the content or formatting of the document. These commands are tracking users' selection of objects/text in the document as well as menu items and UI's toolbars. \\
    \hline
    SubCategories & Sample Commands \\
    \hline
    
    \multirow{5}{*}{SelectObject}
    & SelectTable / SelectRow / SelectCell \\
    & SelectAll \\
    & Pointer \\
    & DrawSelectNext \\
    & SelectionMode \\
    \hline
    \multirow{5}{*}{SelectTab}
    & RefPaneSelection \\
    & TabReview / TabHome \\
    & MergeToolbar \\
    & EndOfLineExtend \\
    & EquationScriptGallery \\
    \hline
    
    \rowcolor[gray]{0.9}Sharing & Commands in this category capture different interactions with sharing the created document either via email or through writing application's built-in sharing features. \\
    \hline
    SubCategories & Sample Commands \\
    \hline
    
    \multirow{5}{*}{Email}
    & MergeToEMail \\
    & Envelope / EnvelopeSend \\
    & TabMailing \\
    & MailMergeCreateList \\
    & AddressBook \\
    \hline
    \multirow{4}{*}{Share}
    & Share \\
    & FileShareMenu \\
    & SendForReview \\
    & SharedWithPersonaCanView \\
    \hline
    
    \rowcolor[gray]{0.9}Communicating & Commands in this category capture different ways of collaboration and communication between co-authors of a shared document, either through comment based communication or tracking and responding to changes. \\
    \hline
    SubCategories & Sample Commands \\
    \hline
    
    \multirow{4}{*}{Comments}
    & NewComment \\
    & DeleteComment \\
    & ReplyToComment \\
    & MarkCommentDone \\
    \hline
    \multirow{5}{*}{ChangeTracking}
    & NextRevision \\
    & AcceptRevision \\
    & RejectAllChangesShown \\
    & RevisionsHighlightChanges \\
    & AcceptConflict \\
    \hline
    
    \rowcolor[gray]{0.9}Finalizing & Commands in this category are mainly applied to a completed document or as preparation for sharing or publishing. These commands include adding restrictions to the document or the audience who can access the file as well as different inspections that can ensure the document is now ready for sharing or publishing. \\
    \hline
    SubCategories & Sample Commands \\
    \hline
    
    \multirow{4}{*}{Publish}
    & DocEncryption \\
    & MarkAsReadOnly \\
    & FileFinalizeMenu \\
    & RestrictFormatting \\
    \hline
    \multirow{5}{*}{Inspecting}
    & WordCompatChkr \\
    & StatusHasPolicy \\
    & DocInspector \\
    & AccessibilityChecker \\
    & DocProtectionManage \\
    \hline
    
    \rowcolor[gray]{0.9}Start / New / Open / Close & These commands simply indicate a new document is created or the writing tool is launched. \\
    \hline
    SubCategories & Sample Commands \\
    \hline
    
    \multirow{1}{*}{Start}
    & OfficeStart \\
    \hline
    \multirow{3}{*}{New}
    & FileNewDialog \\
    & New \\
    & NewDefault \\
    \hline
    Open / Close & These commands simply indicate a document is opened or closed through the menu. \\
    \hline
    
    \rowcolor[gray]{0.9}Save / Print / Copy & These commands simply indicate the user has saved or printed the document or copied some content from this document. These commands could potentially indicate some stage of (partial) completion during the document's lifetime. \\
    \hline
    SubCategories & Sample Commands \\
    \hline
    
    \multirow{4}{*}{Save}
    & Save / SaveAs / SaveAll \\
    & SaveToWebSignIn \\
    & DocExport \\
    & ConvertDoc \\
    \hline
    \multirow{4}{*}{Print}
    & PrintPreviewAndPrint \\
    & Print \\
    & FileNewPrint \\
    & ClosePreview \\
    \hline
    \multirow{3}{*}{Copy}
    & Copy \\
    & CopyText \\
    & WebCopyHyperlink \\
    \hline
    
    \rowcolor[gray]{0.9}Meta & These commands include different functionalities of a modern writing environment that is beyond document creation, authoring or viewing. \\
    \hline
    SubCategories & Sample Commands \\
    \hline
    
    \multirow{5}{*}{Settings}
    & StatusBarConfigMenu \\
    & Custom \\
    & TrackChangesAdvancedOptions \\
    & Ruler11 \\
    & EquationLinearAll \\
    \hline
    \multirow{4}{*}{Help}
    & TabHelp \\
    & Insights \\
    & HelpContactUs \\
    & WhatsNewRecentUpdates \\
    \hline
    \multirow{3}{*}{DocRecovery}
    & StatusDocRecovery \\
    & ShowRepairs \\
    & OpenDrp / ViewDrp \\
    \hline
    \multirow{4}{*}{Macros}
    & StatusRecordMacro \\
    & PauseRecorder \\
    & TabDeveloperTools \\
    & Play \\
    \hline
    \multirow{3}{*}{AddIns}
    & OpenLinkedIn \\
    & OfficeExtensionsGallery \\
    & TabAddins \\
    \hline
\end{tabularx}
\newpage
\section{Overview of Features Used for Prediction}
\label{Appendix_features}
\begin{table}[hb]
\caption{Feature Categories used for temporal stage of document evolution prediction.}
\begin{small}
\begin{tabularx}{\textwidth}{m{3cm}|X}
  \hline
  \multicolumn{1}{c}{\textbf{Name}} & \multicolumn{1}{c}{\textbf{Description}}\\
  \hline
  \rowcolor[gray]{.9}\multicolumn{2}{l}{Command Features} \\
  \hline
  See Appendix A & The lowest level of taxonomy described in Section \ref{LogDataAnalysis}\\
  \hline
  \rowcolor[gray]{.9}\multicolumn{2}{l}{Command Category Features} \\
  \hline
  See Appendix A & The second level of taxonomy described in Section \ref{LogDataAnalysis} \\
  \hline
  \rowcolor[gray]{.9}\multicolumn{2}{l}{High Level Command Category Features} \\
  \hline
  See Appendix A & The highest level of taxonomy described in Section \ref{LogDataAnalysis} \\
  \hline
  \rowcolor[gray]{.9}\multicolumn{2}{l}{Advanced Functionality Features} \\
  \hline
  BulletsAndNumbering & num. lists + whether there is a list (i.e. binary feature) \\
  Canvas & whether the drawing canvas is used  \\ 
  HeaderFooter & whether HeaderFooter is enabled \\
  Shapes & num. shapes + whether there is a shape (i.e. binary feature) \\
  Bibliography & whether Bibliography is inserted \\
  Hyperlinks & num. hyperlinks + whether there is a hyperlink (i.e. binary feature) \\
  ... & These features correspond to the analysis discussed in Section \ref{Logs_TypesOfActivities} \\
  \hline
  \rowcolor[gray]{.9}\multicolumn{2}{l}{Collaboration Features} \\
  \hline
  \# Collaborators & Collaborators are users who contributed to the document in at least one of adding, editing, communicating or finalizing \\
  CBS &  Contribution Balance Score reflects how balanced the contribution was across all collaborators regardless of the type of contribution.\\
  RBS-adding & Whether all collaborators contributed equally to adding content\\
  RBS-editing & Whether all collaborators contributed equally to editing\\
  RBS-communicating & Whether all collaborators contributed equally to communicating\\
  RBS-finalizing & Whether all collaborators contributed equally to finalizing\\
  ... & RBS scores can be calculated for each of command categories using the formula introduced in Section \ref{predictionSection}\\
\hline
\rowcolor[gray]{.9}\multicolumn{2}{l}{Content Features} \\
  \hline
  PageCount & num. pages in the document at a given timestamp \\
  SectionsCount & num. sections in the document at a given timestamp  \\ 
  ParagraphCount & num. paragraphs in the document at a given timestamp \\
  LineCount & num. lines in the document at a given timestamp \\
  WordCount & num. words in the document at a given timestamp \\
  CharacterCount & num. characters in the document at a given timestamp \\
  \hline
\rowcolor[gray]{.9}\multicolumn{2}{l}{Lifetime Features} \\
  \hline
  TimeElapsed & amount of time in milliseconds that has passed since document creation \\
  TimeElapsedBucket & A coarse grained bucketization of TimeElapsed as numDays, numWeeks, etc.  \\ 
  \hline
\end{tabularx}
\end{small}
\label{tab:features}
\end{table}
\end{document}